\documentclass[year=22,pdfa]{fmcad}

\ifCLASSOPTIONcompsoc
  \usepackage[nocompress]{cite}
\else
  \usepackage{cite}
\fi

\usepackage{graphicx}
\usepackage{subcaption}
\usepackage{xspace}
\usepackage{tikz}
\usetikzlibrary{arrows,fit,positioning,shapes}
\usepackage{hyperref}
\usepackage{amsmath}
\usepackage{amsthm}
\usepackage{cleveref}
\usepackage{mathtools}

\usepackage{color}
\usepackage{listings}
\usepackage[textsize=tiny]{todonotes}

\author{\IEEEauthorblockN{Oliver Flatt}
\IEEEauthorblockA{University of Washington\\
Seattle WA 98195, USA\\
Email: oflatt@cs.washington.edu}
\and
\IEEEauthorblockN{Samuel Coward}
\IEEEauthorblockA{Numerical Hardware Group \\
Intel Corporation \\
samuel.coward@intel.com}
\and
\IEEEauthorblockN{Max Willsey}
\IEEEauthorblockA{University of Washington\\
Seattle WA 98195, USA\\
Email: mwillsey@cs.washington.edu}
\and
\IEEEauthorblockN{Zachary Tatlock}
\IEEEauthorblockA{University of Washington\\
Seattle WA 98195, USA\\
Email: ztatlock@cs.washington.edu}
\and
\IEEEauthorblockN{Pavel Panchekha}
\IEEEauthorblockA{University of Utah\\
Salt Lake City, UT 84112, USA\\
Email: pavpan@cs.utah.edu}
}

\begin{document}

\newtheorem{definition}{Definition}
\newtheorem{theorem}{Theorem}

\title{Small Proofs from Congruence Closure}
\maketitle

\newcommand{\proofs}{\mbox{proofs}\xspace}
\newcommand{\eproof}{\mbox{proof}\xspace}
\newcommand{\Proof}{\mbox{Proof}\xspace}
\newcommand{\Proofs}{\mbox{Proofs}\xspace}

\newcommand{\egraphs}{\mbox{e-graphs}\xspace}
\newcommand{\egraph}{\mbox{e-graph}\xspace}
\newcommand{\Egraph}{\mbox{E-graph}\xspace}
\newcommand{\Egraphs}{\mbox{E-graphs}\xspace}
\newcommand{\eclass}{\mbox{e-class}\xspace}
\newcommand{\Eclass}{\mbox{E-class}\xspace}
\newcommand{\enode}{\mbox{e-node}\xspace}
\newcommand{\eclasses}{\mbox{e-classes}\xspace}
\newcommand{\enodes}{\mbox{e-nodes}\xspace}

\newcommand{\cgraphs}{\mbox{c-graphs}\xspace}
\newcommand{\cgraph}{\mbox{c-graph}\xspace}
\newcommand{\Cgraph}{\mbox{C-graph}\xspace}
\newcommand{\Cgraphs}{\mbox{C-graphs}\xspace}

\newcommand{\treeopt}{\mbox{TreeOpt}\xspace}

\newcommand{\egg}{\texorpdfstring{\MakeLowercase{\texttt{egg}}}{\texttt{egg}}\xspace}

\newcommand{\mypara}[1]{\vspace{1ex}\noindent\textit{\textbf{#1}}}
\newcommand{\rrr}{r}

\newcommand{\prooflengthnumbenchmarks}{3\thinspace760\xspace}
\newcommand{\prooflengthvanillasum}{88\thinspace173\xspace}
\newcommand{\prooflengthsum}{36\thinspace682\xspace}
\newcommand{\prooflengthpercentsmallerthanvanilla}{53.1\%\xspace}
\newcommand{\prooflengthpercentsmallerthanhalfvanilla}{24.0\%\xspace}
\newcommand{\prooflengthpercentreduction}{58.4\%\xspace}
\newcommand{\prooflengthpercentasbig}{41.6\%\xspace}
\newcommand{\prooflengthbestpercentasbig}{1.4\%\xspace}
\newcommand{\prooflengthotherpercentasbig}{240.4\%\xspace}

\newcommand{\prooflengthnumbenchmarkslengthgrtten}{1\thinspace732\xspace}
\newcommand{\prooflengthvanillasumlengthgrtten}{78\thinspace788\xspace}
\newcommand{\prooflengthsumlengthgrtten}{28\thinspace524\xspace}
\newcommand{\prooflengthpercentsmallerthanvanillalengthgrtten}{88.2\%\xspace}
\newcommand{\prooflengthpercentsmallerthanhalfvanillalengthgrtten}{47.9\%\xspace}
\newcommand{\prooflengthpercentreductionlengthgrtten}{63.8\%\xspace}
\newcommand{\prooflengthpercentasbiglengthgrtten}{36.2\%\xspace}
\newcommand{\prooflengthbestpercentasbiglengthgrtten}{1.4\%\xspace}
\newcommand{\prooflengthotherpercentasbiglengthgrtten}{276.2\%\xspace}

\newcommand{\prooflengthnumbenchmarkslengthgrtfifty}{399\xspace}
\newcommand{\prooflengthvanillasumlengthgrtfifty}{47\thinspace648\xspace}
\newcommand{\prooflengthsumlengthgrtfifty}{10\thinspace884\xspace}
\newcommand{\prooflengthpercentsmallerthanvanillalengthgrtfifty}{98.7\%\xspace}
\newcommand{\prooflengthpercentsmallerthanhalfvanillalengthgrtfifty}{85.7\%\xspace}
\newcommand{\prooflengthpercentreductionlengthgrtfifty}{77.2\%\xspace}
\newcommand{\prooflengthpercentasbiglengthgrtfifty}{22.8\%\xspace}
\newcommand{\prooflengthbestpercentasbiglengthgrtfifty}{1.4\%\xspace}
\newcommand{\prooflengthotherpercentasbiglengthgrtfifty}{437.8\%\xspace}

\newcommand{\dagsizenumbenchmarks}{3\thinspace760\xspace}
\newcommand{\dagsizevanillasum}{48\thinspace233\xspace}
\newcommand{\dagsizesum}{26\thinspace841\xspace}
\newcommand{\dagsizepercentsmallerthanvanilla}{51.4\%\xspace}
\newcommand{\dagsizepercentsmallerthanhalfvanilla}{20.9\%\xspace}
\newcommand{\dagsizepercentreduction}{44.4\%\xspace}
\newcommand{\dagsizepercentasbig}{55.6\%\xspace}
\newcommand{\dagsizebestpercentasbig}{2.3\%\xspace}
\newcommand{\dagsizeotherpercentasbig}{179.7\%\xspace}

\newcommand{\dagsizenumbenchmarkslengthgrtten}{1\thinspace430\xspace}
\newcommand{\dagsizevanillasumlengthgrtten}{38\thinspace805\xspace}
\newcommand{\dagsizesumlengthgrtten}{19\thinspace139\xspace}
\newcommand{\dagsizepercentsmallerthanvanillalengthgrtten}{89.0\%\xspace}
\newcommand{\dagsizepercentsmallerthanhalfvanillalengthgrtten}{42.9\%\xspace}
\newcommand{\dagsizepercentreductionlengthgrtten}{50.7\%\xspace}
\newcommand{\dagsizepercentasbiglengthgrtten}{49.3\%\xspace}
\newcommand{\dagsizebestpercentasbiglengthgrtten}{2.3\%\xspace}
\newcommand{\dagsizeotherpercentasbiglengthgrtten}{202.8\%\xspace}

\newcommand{\dagsizenumbenchmarkslengthgrtfifty}{148\xspace}
\newcommand{\dagsizevanillasumlengthgrtfifty}{10\thinspace094\xspace}
\newcommand{\dagsizesumlengthgrtfifty}{3\thinspace474\xspace}
\newcommand{\dagsizepercentsmallerthanvanillalengthgrtfifty}{98.6\%\xspace}
\newcommand{\dagsizepercentsmallerthanhalfvanillalengthgrtfifty}{82.4\%\xspace}
\newcommand{\dagsizepercentreductionlengthgrtfifty}{65.6\%\xspace}
\newcommand{\dagsizepercentasbiglengthgrtfifty}{34.4\%\xspace}
\newcommand{\dagsizebestpercentasbiglengthgrtfifty}{2.3\%\xspace}
\newcommand{\dagsizeotherpercentasbiglengthgrtfifty}{290.6\%\xspace}

\newcommand{\zdagsizenumbenchmarks}{3\thinspace571\xspace}
\newcommand{\zdagsizevanillasum}{32\thinspace910\xspace}
\newcommand{\zdagsizesum}{23\thinspace968\xspace}
\newcommand{\zdagsizepercentsmallerthanvanilla}{44.1\%\xspace}
\newcommand{\zdagsizepercentsmallerthanhalfvanilla}{7.4\%\xspace}
\newcommand{\zdagsizepercentreduction}{27.2\%\xspace}
\newcommand{\zdagsizepercentasbig}{72.8\%\xspace}
\newcommand{\zdagsizebestpercentasbig}{0.0\%\xspace}
\newcommand{\zdagsizeotherpercentasbig}{137.3\%\xspace}

\newcommand{\zdagsizenumbenchmarkslengthgrtten}{1\thinspace036\xspace}
\newcommand{\zdagsizevanillasumlengthgrtten}{23\thinspace324\xspace}
\newcommand{\zdagsizesumlengthgrtten}{14\thinspace930\xspace}
\newcommand{\zdagsizepercentsmallerthanvanillalengthgrtten}{87.0\%\xspace}
\newcommand{\zdagsizepercentsmallerthanhalfvanillalengthgrtten}{20.1\%\xspace}
\newcommand{\zdagsizepercentreductionlengthgrtten}{36.0\%\xspace}
\newcommand{\zdagsizepercentasbiglengthgrtten}{64.0\%\xspace}
\newcommand{\zdagsizebestpercentasbiglengthgrtten}{15.0\%\xspace}
\newcommand{\zdagsizeotherpercentasbiglengthgrtten}{156.2\%\xspace}

\newcommand{\zdagsizenumbenchmarkslengthgrtfifty}{36\xspace}
\newcommand{\zdagsizevanillasumlengthgrtfifty}{2\thinspace373\xspace}
\newcommand{\zdagsizesumlengthgrtfifty}{1\thinspace193\xspace}
\newcommand{\zdagsizepercentsmallerthanvanillalengthgrtfifty}{100.0\%\xspace}
\newcommand{\zdagsizepercentsmallerthanhalfvanillalengthgrtfifty}{50.0\%\xspace}
\newcommand{\zdagsizepercentreductionlengthgrtfifty}{49.7\%\xspace}
\newcommand{\zdagsizepercentasbiglengthgrtfifty}{50.3\%\xspace}
\newcommand{\zdagsizebestpercentasbiglengthgrtfifty}{22.9\%\xspace}
\newcommand{\zdagsizeotherpercentasbiglengthgrtfifty}{198.9\%\xspace}

\newcommand{\upwardsdagsizenumbenchmarks}{3\thinspace205\xspace}
\newcommand{\upwardsdagsizevanillasum}{39\thinspace350\xspace}
\newcommand{\upwardsdagsizesum}{51\thinspace491\xspace}
\newcommand{\upwardsdagsizepercentsmallerthanvanilla}{29.5\%\xspace}
\newcommand{\upwardsdagsizepercentsmallerthanhalfvanilla}{10.6\%\xspace}
\newcommand{\upwardsdagsizepercentreduction}{-30.9\%\xspace}
\newcommand{\upwardsdagsizepercentasbig}{130.9\%\xspace}
\newcommand{\upwardsdagsizebestpercentasbig}{2.3\%\xspace}
\newcommand{\upwardsdagsizeotherpercentasbig}{76.4\%\xspace}

\newcommand{\upwardsdagsizenumbenchmarkslengthgrtten}{1\thinspace137\xspace}
\newcommand{\upwardsdagsizevanillasumlengthgrtten}{31\thinspace261\xspace}
\newcommand{\upwardsdagsizesumlengthgrtten}{36\thinspace907\xspace}
\newcommand{\upwardsdagsizepercentsmallerthanvanillalengthgrtten}{53.6\%\xspace}
\newcommand{\upwardsdagsizepercentsmallerthanhalfvanillalengthgrtten}{23.1\%\xspace}
\newcommand{\upwardsdagsizepercentreductionlengthgrtten}{-18.1\%\xspace}
\newcommand{\upwardsdagsizepercentasbiglengthgrtten}{118.1\%\xspace}
\newcommand{\upwardsdagsizebestpercentasbiglengthgrtten}{2.3\%\xspace}
\newcommand{\upwardsdagsizeotherpercentasbiglengthgrtten}{84.7\%\xspace}

\newcommand{\upwardsdagsizenumbenchmarkslengthgrtfifty}{122\xspace}
\newcommand{\upwardsdagsizevanillasumlengthgrtfifty}{8\thinspace274\xspace}
\newcommand{\upwardsdagsizesumlengthgrtfifty}{7\thinspace011\xspace}
\newcommand{\upwardsdagsizepercentsmallerthanvanillalengthgrtfifty}{72.1\%\xspace}
\newcommand{\upwardsdagsizepercentsmallerthanhalfvanillalengthgrtfifty}{42.6\%\xspace}
\newcommand{\upwardsdagsizepercentreductionlengthgrtfifty}{15.3\%\xspace}
\newcommand{\upwardsdagsizepercentasbiglengthgrtfifty}{84.7\%\xspace}
\newcommand{\upwardsdagsizebestpercentasbiglengthgrtfifty}{2.3\%\xspace}
\newcommand{\upwardsdagsizeotherpercentasbiglengthgrtfifty}{118.0\%\xspace}

\newcommand{\upwardsvszdagsizenumbenchmarks}{3\thinspace058\xspace}
\newcommand{\upwardsvszdagsizevanillasum}{25\thinspace723\xspace}
\newcommand{\upwardsvszdagsizesum}{45\thinspace013\xspace}
\newcommand{\upwardsvszdagsizepercentsmallerthanvanilla}{20.6\%\xspace}
\newcommand{\upwardsvszdagsizepercentsmallerthanhalfvanilla}{1.8\%\xspace}
\newcommand{\upwardsvszdagsizepercentreduction}{-75.0\%\xspace}
\newcommand{\upwardsvszdagsizepercentasbig}{175.0\%\xspace}
\newcommand{\upwardsvszdagsizebestpercentasbig}{0.0\%\xspace}
\newcommand{\upwardsvszdagsizeotherpercentasbig}{57.1\%\xspace}

\newcommand{\upwardsvszdagsizenumbenchmarkslengthgrtten}{800\xspace}
\newcommand{\upwardsvszdagsizevanillasumlengthgrtten}{17\thinspace383\xspace}
\newcommand{\upwardsvszdagsizesumlengthgrtten}{27\thinspace125\xspace}
\newcommand{\upwardsvszdagsizepercentsmallerthanvanillalengthgrtten}{38.0\%\xspace}
\newcommand{\upwardsvszdagsizepercentsmallerthanhalfvanillalengthgrtten}{3.5\%\xspace}
\newcommand{\upwardsvszdagsizepercentreductionlengthgrtten}{-56.0\%\xspace}
\newcommand{\upwardsvszdagsizepercentasbiglengthgrtten}{156.0\%\xspace}
\newcommand{\upwardsvszdagsizebestpercentasbiglengthgrtten}{20.0\%\xspace}
\newcommand{\upwardsvszdagsizeotherpercentasbiglengthgrtten}{64.1\%\xspace}

\newcommand{\upwardsvszdagsizenumbenchmarkslengthgrtfifty}{22\xspace}
\newcommand{\upwardsvszdagsizevanillasumlengthgrtfifty}{1\thinspace488\xspace}
\newcommand{\upwardsvszdagsizesumlengthgrtfifty}{1\thinspace922\xspace}
\newcommand{\upwardsvszdagsizepercentsmallerthanvanillalengthgrtfifty}{40.9\%\xspace}
\newcommand{\upwardsvszdagsizepercentsmallerthanhalfvanillalengthgrtfifty}{9.1\%\xspace}
\newcommand{\upwardsvszdagsizepercentreductionlengthgrtfifty}{-29.2\%\xspace}
\newcommand{\upwardsvszdagsizepercentasbiglengthgrtfifty}{129.2\%\xspace}
\newcommand{\upwardsvszdagsizebestpercentasbiglengthgrtfifty}{33.9\%\xspace}
\newcommand{\upwardsvszdagsizeotherpercentasbiglengthgrtfifty}{77.4\%\xspace}

\newcommand{\lowtreesizenumbenchmarks}{3\thinspace760\xspace}
\newcommand{\lowtreesizevanillasum}{36\thinspace682\xspace}
\newcommand{\lowtreesizesum}{33\thinspace677\xspace}
\newcommand{\lowtreesizepercentsmallerthanvanilla}{17.8\%\xspace}
\newcommand{\lowtreesizepercentsmallerthanhalfvanilla}{1.2\%\xspace}
\newcommand{\lowtreesizepercentreduction}{8.2\%\xspace}
\newcommand{\lowtreesizepercentasbig}{91.8\%\xspace}
\newcommand{\lowtreesizebestpercentasbig}{8.3\%\xspace}
\newcommand{\lowtreesizeotherpercentasbig}{108.9\%\xspace}

\newcommand{\lowtreesizenumbenchmarkslengthgrtten}{1\thinspace134\xspace}
\newcommand{\lowtreesizevanillasumlengthgrtten}{24\thinspace394\xspace}
\newcommand{\lowtreesizesumlengthgrtten}{21\thinspace700\xspace}
\newcommand{\lowtreesizepercentsmallerthanvanillalengthgrtten}{45.1\%\xspace}
\newcommand{\lowtreesizepercentsmallerthanhalfvanillalengthgrtten}{3.3\%\xspace}
\newcommand{\lowtreesizepercentreductionlengthgrtten}{11.0\%\xspace}
\newcommand{\lowtreesizepercentasbiglengthgrtten}{89.0\%\xspace}
\newcommand{\lowtreesizebestpercentasbiglengthgrtten}{8.3\%\xspace}
\newcommand{\lowtreesizeotherpercentasbiglengthgrtten}{112.4\%\xspace}

\newcommand{\lowtreesizenumbenchmarkslengthgrtfifty}{39\xspace}
\newcommand{\lowtreesizevanillasumlengthgrtfifty}{3\thinspace724\xspace}
\newcommand{\lowtreesizesumlengthgrtfifty}{3\thinspace498\xspace}
\newcommand{\lowtreesizepercentsmallerthanvanillalengthgrtfifty}{53.8\%\xspace}
\newcommand{\lowtreesizepercentsmallerthanhalfvanillalengthgrtfifty}{2.6\%\xspace}
\newcommand{\lowtreesizepercentreductionlengthgrtfifty}{6.1\%\xspace}
\newcommand{\lowtreesizepercentasbiglengthgrtfifty}{93.9\%\xspace}
\newcommand{\lowtreesizebestpercentasbiglengthgrtfifty}{35.8\%\xspace}
\newcommand{\lowtreesizeotherpercentasbiglengthgrtfifty}{106.5\%\xspace}

\newcommand{\reductionvsvanillanumbenchmarks}{3\thinspace760\xspace}
\newcommand{\reductionvsvanillavanillasum}{48\thinspace233\xspace}
\newcommand{\reductionvsvanillasum}{34\thinspace630\xspace}
\newcommand{\reductionvsvanillapercentsmallerthanvanilla}{36.4\%\xspace}
\newcommand{\reductionvsvanillapercentsmallerthanhalfvanilla}{11.3\%\xspace}
\newcommand{\reductionvsvanillapercentreduction}{28.2\%\xspace}
\newcommand{\reductionvsvanillapercentasbig}{71.8\%\xspace}
\newcommand{\reductionvsvanillabestpercentasbig}{2.3\%\xspace}
\newcommand{\reductionvsvanillaotherpercentasbig}{139.3\%\xspace}

\newcommand{\reductionvsvanillanumbenchmarkslengthgrtten}{1\thinspace430\xspace}
\newcommand{\reductionvsvanillavanillasumlengthgrtten}{38\thinspace805\xspace}
\newcommand{\reductionvsvanillasumlengthgrtten}{26\thinspace275\xspace}
\newcommand{\reductionvsvanillapercentsmallerthanvanillalengthgrtten}{70.2\%\xspace}
\newcommand{\reductionvsvanillapercentsmallerthanhalfvanillalengthgrtten}{21.0\%\xspace}
\newcommand{\reductionvsvanillapercentreductionlengthgrtten}{32.3\%\xspace}
\newcommand{\reductionvsvanillapercentasbiglengthgrtten}{67.7\%\xspace}
\newcommand{\reductionvsvanillabestpercentasbiglengthgrtten}{2.3\%\xspace}
\newcommand{\reductionvsvanillaotherpercentasbiglengthgrtten}{147.7\%\xspace}

\newcommand{\reductionvsvanillanumbenchmarkslengthgrtfifty}{148\xspace}
\newcommand{\reductionvsvanillavanillasumlengthgrtfifty}{10\thinspace094\xspace}
\newcommand{\reductionvsvanillasumlengthgrtfifty}{5\thinspace776\xspace}
\newcommand{\reductionvsvanillapercentsmallerthanvanillalengthgrtfifty}{96.6\%\xspace}
\newcommand{\reductionvsvanillapercentsmallerthanhalfvanillalengthgrtfifty}{41.2\%\xspace}
\newcommand{\reductionvsvanillapercentreductionlengthgrtfifty}{42.8\%\xspace}
\newcommand{\reductionvsvanillapercentasbiglengthgrtfifty}{57.2\%\xspace}
\newcommand{\reductionvsvanillabestpercentasbiglengthgrtfifty}{2.3\%\xspace}
\newcommand{\reductionvsvanillaotherpercentasbiglengthgrtfifty}{174.8\%\xspace}

\newcommand{\greedyandreductionvsvanillanumbenchmarks}{3\thinspace760\xspace}
\newcommand{\greedyandreductionvsvanillavanillasum}{48\thinspace233\xspace}
\newcommand{\greedyandreductionvsvanillasum}{26\thinspace188\xspace}
\newcommand{\greedyandreductionvsvanillapercentsmallerthanvanilla}{51.8\%\xspace}
\newcommand{\greedyandreductionvsvanillapercentsmallerthanhalfvanilla}{22.3\%\xspace}
\newcommand{\greedyandreductionvsvanillapercentreduction}{45.7\%\xspace}
\newcommand{\greedyandreductionvsvanillapercentasbig}{54.3\%\xspace}
\newcommand{\greedyandreductionvsvanillabestpercentasbig}{2.3\%\xspace}
\newcommand{\greedyandreductionvsvanillaotherpercentasbig}{184.2\%\xspace}

\newcommand{\greedyandreductionvsvanillanumbenchmarkslengthgrtten}{1\thinspace430\xspace}
\newcommand{\greedyandreductionvsvanillavanillasumlengthgrtten}{38\thinspace805\xspace}
\newcommand{\greedyandreductionvsvanillasumlengthgrtten}{18\thinspace538\xspace}
\newcommand{\greedyandreductionvsvanillapercentsmallerthanvanillalengthgrtten}{89.0\%\xspace}
\newcommand{\greedyandreductionvsvanillapercentsmallerthanhalfvanillalengthgrtten}{46.1\%\xspace}
\newcommand{\greedyandreductionvsvanillapercentreductionlengthgrtten}{52.2\%\xspace}
\newcommand{\greedyandreductionvsvanillapercentasbiglengthgrtten}{47.8\%\xspace}
\newcommand{\greedyandreductionvsvanillabestpercentasbiglengthgrtten}{2.3\%\xspace}
\newcommand{\greedyandreductionvsvanillaotherpercentasbiglengthgrtten}{209.3\%\xspace}

\newcommand{\greedyandreductionvsvanillanumbenchmarkslengthgrtfifty}{148\xspace}
\newcommand{\greedyandreductionvsvanillavanillasumlengthgrtfifty}{10\thinspace094\xspace}
\newcommand{\greedyandreductionvsvanillasumlengthgrtfifty}{3\thinspace367\xspace}
\newcommand{\greedyandreductionvsvanillapercentsmallerthanvanillalengthgrtfifty}{98.6\%\xspace}
\newcommand{\greedyandreductionvsvanillapercentsmallerthanhalfvanillalengthgrtfifty}{83.1\%\xspace}
\newcommand{\greedyandreductionvsvanillapercentreductionlengthgrtfifty}{66.6\%\xspace}
\newcommand{\greedyandreductionvsvanillapercentasbiglengthgrtfifty}{33.4\%\xspace}
\newcommand{\greedyandreductionvsvanillabestpercentasbiglengthgrtfifty}{2.3\%\xspace}
\newcommand{\greedyandreductionvsvanillaotherpercentasbiglengthgrtfifty}{299.8\%\xspace}

\newcommand{\greedyvsreductionnumbenchmarks}{3\thinspace760\xspace}
\newcommand{\greedyvsreductionvanillasum}{34\thinspace630\xspace}
\newcommand{\greedyvsreductionsum}{26\thinspace841\xspace}
\newcommand{\greedyvsreductionpercentsmallerthanvanilla}{34.9\%\xspace}
\newcommand{\greedyvsreductionpercentsmallerthanhalfvanilla}{6.5\%\xspace}
\newcommand{\greedyvsreductionpercentreduction}{22.5\%\xspace}
\newcommand{\greedyvsreductionpercentasbig}{77.5\%\xspace}
\newcommand{\greedyvsreductionbestpercentasbig}{5.1\%\xspace}
\newcommand{\greedyvsreductionotherpercentasbig}{129.0\%\xspace}

\newcommand{\greedyvsreductionnumbenchmarkslengthgrtten}{1\thinspace118\xspace}
\newcommand{\greedyvsreductionvanillasumlengthgrtten}{24\thinspace274\xspace}
\newcommand{\greedyvsreductionsumlengthgrtten}{17\thinspace126\xspace}
\newcommand{\greedyvsreductionpercentsmallerthanvanillalengthgrtten}{74.5\%\xspace}
\newcommand{\greedyvsreductionpercentsmallerthanhalfvanillalengthgrtten}{17.8\%\xspace}
\newcommand{\greedyvsreductionpercentreductionlengthgrtten}{29.4\%\xspace}
\newcommand{\greedyvsreductionpercentasbiglengthgrtten}{70.6\%\xspace}
\newcommand{\greedyvsreductionbestpercentasbiglengthgrtten}{5.1\%\xspace}
\newcommand{\greedyvsreductionotherpercentasbiglengthgrtten}{141.7\%\xspace}

\newcommand{\greedyvsreductionnumbenchmarkslengthgrtfifty}{33\xspace}
\newcommand{\greedyvsreductionvanillasumlengthgrtfifty}{2\thinspace245\xspace}
\newcommand{\greedyvsreductionsumlengthgrtfifty}{1\thinspace295\xspace}
\newcommand{\greedyvsreductionpercentsmallerthanvanillalengthgrtfifty}{78.8\%\xspace}
\newcommand{\greedyvsreductionpercentsmallerthanhalfvanillalengthgrtfifty}{54.5\%\xspace}
\newcommand{\greedyvsreductionpercentreductionlengthgrtfifty}{42.3\%\xspace}
\newcommand{\greedyvsreductionpercentasbiglengthgrtfifty}{57.7\%\xspace}
\newcommand{\greedyvsreductionbestpercentasbiglengthgrtfifty}{20.0\%\xspace}
\newcommand{\greedyvsreductionotherpercentasbiglengthgrtfifty}{173.4\%\xspace}

\newcommand{\eqcheckdagsizevsvanillanumbenchmarks}{3\thinspace752\xspace}
\newcommand{\eqcheckdagsizevsvanillavanillasum}{43\thinspace661\xspace}
\newcommand{\eqcheckdagsizevsvanillasum}{47\thinspace919\xspace}
\newcommand{\eqcheckdagsizevsvanillapercentsmallerthanvanilla}{27.3\%\xspace}
\newcommand{\eqcheckdagsizevsvanillapercentsmallerthanhalfvanilla}{5.0\%\xspace}
\newcommand{\eqcheckdagsizevsvanillapercentreduction}{-9.8\%\xspace}
\newcommand{\eqcheckdagsizevsvanillapercentasbig}{109.8\%\xspace}
\newcommand{\eqcheckdagsizevsvanillabestpercentasbig}{10.5\%\xspace}
\newcommand{\eqcheckdagsizevsvanillaotherpercentasbig}{91.1\%\xspace}

\newcommand{\eqcheckdagsizevsvanillanumbenchmarkslengthgrtten}{1\thinspace318\xspace}
\newcommand{\eqcheckdagsizevsvanillavanillasumlengthgrtten}{33\thinspace975\xspace}
\newcommand{\eqcheckdagsizevsvanillasumlengthgrtten}{33\thinspace745\xspace}
\newcommand{\eqcheckdagsizevsvanillapercentsmallerthanvanillalengthgrtten}{51.3\%\xspace}
\newcommand{\eqcheckdagsizevsvanillapercentsmallerthanhalfvanillalengthgrtten}{9.0\%\xspace}
\newcommand{\eqcheckdagsizevsvanillapercentreductionlengthgrtten}{0.7\%\xspace}
\newcommand{\eqcheckdagsizevsvanillapercentasbiglengthgrtten}{99.3\%\xspace}
\newcommand{\eqcheckdagsizevsvanillabestpercentasbiglengthgrtten}{10.5\%\xspace}
\newcommand{\eqcheckdagsizevsvanillaotherpercentasbiglengthgrtten}{100.7\%\xspace}

\newcommand{\eqcheckdagsizevsvanillanumbenchmarkslengthgrtfifty}{107\xspace}
\newcommand{\eqcheckdagsizevsvanillavanillasumlengthgrtfifty}{7\thinspace843\xspace}
\newcommand{\eqcheckdagsizevsvanillasumlengthgrtfifty}{5\thinspace725\xspace}
\newcommand{\eqcheckdagsizevsvanillapercentsmallerthanvanillalengthgrtfifty}{79.4\%\xspace}
\newcommand{\eqcheckdagsizevsvanillapercentsmallerthanhalfvanillalengthgrtfifty}{23.4\%\xspace}
\newcommand{\eqcheckdagsizevsvanillapercentreductionlengthgrtfifty}{27.0\%\xspace}
\newcommand{\eqcheckdagsizevsvanillapercentasbiglengthgrtfifty}{73.0\%\xspace}
\newcommand{\eqcheckdagsizevsvanillabestpercentasbiglengthgrtfifty}{15.1\%\xspace}
\newcommand{\eqcheckdagsizevsvanillaotherpercentasbiglengthgrtfifty}{137.0\%\xspace}

\newcommand{\optimaldagsizevsgreedynumbenchmarks}{3\thinspace760\xspace}
\newcommand{\optimaldagsizevsgreedyvanillasum}{26\thinspace841\xspace}
\newcommand{\optimaldagsizevsgreedysum}{24\thinspace899\xspace}
\newcommand{\optimaldagsizevsgreedypercentsmallerthanvanilla}{17.3\%\xspace}
\newcommand{\optimaldagsizevsgreedypercentsmallerthanhalfvanilla}{1.0\%\xspace}
\newcommand{\optimaldagsizevsgreedypercentreduction}{7.2\%\xspace}
\newcommand{\optimaldagsizevsgreedypercentasbig}{92.8\%\xspace}
\newcommand{\optimaldagsizevsgreedybestpercentasbig}{10.0\%\xspace}
\newcommand{\optimaldagsizevsgreedyotherpercentasbig}{107.8\%\xspace}

\newcommand{\optimaldagsizevsgreedynumbenchmarkslengthgrtten}{858\xspace}
\newcommand{\optimaldagsizevsgreedyvanillasumlengthgrtten}{15\thinspace300\xspace}
\newcommand{\optimaldagsizevsgreedysumlengthgrtten}{13\thinspace752\xspace}
\newcommand{\optimaldagsizevsgreedypercentsmallerthanvanillalengthgrtten}{47.4\%\xspace}
\newcommand{\optimaldagsizevsgreedypercentsmallerthanhalfvanillalengthgrtten}{2.7\%\xspace}
\newcommand{\optimaldagsizevsgreedypercentreductionlengthgrtten}{10.1\%\xspace}
\newcommand{\optimaldagsizevsgreedypercentasbiglengthgrtten}{89.9\%\xspace}
\newcommand{\optimaldagsizevsgreedybestpercentasbiglengthgrtten}{11.1\%\xspace}
\newcommand{\optimaldagsizevsgreedyotherpercentasbiglengthgrtten}{111.3\%\xspace}

\newcommand{\optimaldagsizevsgreedynumbenchmarkslengthgrtfifty}{12\xspace}
\newcommand{\optimaldagsizevsgreedyvanillasumlengthgrtfifty}{812\xspace}
\newcommand{\optimaldagsizevsgreedysumlengthgrtfifty}{794\xspace}
\newcommand{\optimaldagsizevsgreedypercentsmallerthanvanillalengthgrtfifty}{41.7\%\xspace}
\newcommand{\optimaldagsizevsgreedypercentsmallerthanhalfvanillalengthgrtfifty}{0.0\%\xspace}
\newcommand{\optimaldagsizevsgreedypercentreductionlengthgrtfifty}{2.2\%\xspace}
\newcommand{\optimaldagsizevsgreedypercentasbiglengthgrtfifty}{97.8\%\xspace}
\newcommand{\optimaldagsizevsgreedybestpercentasbiglengthgrtfifty}{91.4\%\xspace}
\newcommand{\optimaldagsizevsgreedyotherpercentasbiglengthgrtfifty}{102.3\%\xspace}

\newcommand{\zdagsizevsoptimalnumbenchmarks}{3\thinspace571\xspace}
\newcommand{\zdagsizevsoptimalvanillasum}{22\thinspace299\xspace}
\newcommand{\zdagsizevsoptimalsum}{32\thinspace910\xspace}
\newcommand{\zdagsizevsoptimalpercentsmallerthanvanilla}{7.5\%\xspace}
\newcommand{\zdagsizevsoptimalpercentsmallerthanhalfvanilla}{0.5\%\xspace}
\newcommand{\zdagsizevsoptimalpercentreduction}{-47.6\%\xspace}
\newcommand{\zdagsizevsoptimalpercentasbig}{147.6\%\xspace}
\newcommand{\zdagsizevsoptimalbestpercentasbig}{0.0\%\xspace}
\newcommand{\zdagsizevsoptimalotherpercentasbig}{67.8\%\xspace}

\newcommand{\zdagsizevsoptimalnumbenchmarkslengthgrtten}{669\xspace}
\newcommand{\zdagsizevsoptimalvanillasumlengthgrtten}{10\thinspace987\xspace}
\newcommand{\zdagsizevsoptimalsumlengthgrtten}{17\thinspace505\xspace}
\newcommand{\zdagsizevsoptimalpercentsmallerthanvanillalengthgrtten}{8.5\%\xspace}
\newcommand{\zdagsizevsoptimalpercentsmallerthanhalfvanillalengthgrtten}{0.4\%\xspace}
\newcommand{\zdagsizevsoptimalpercentreductionlengthgrtten}{-59.3\%\xspace}
\newcommand{\zdagsizevsoptimalpercentasbiglengthgrtten}{159.3\%\xspace}
\newcommand{\zdagsizevsoptimalbestpercentasbiglengthgrtten}{33.3\%\xspace}
\newcommand{\zdagsizevsoptimalotherpercentasbiglengthgrtten}{62.8\%\xspace}

\newcommand{\zdagsizevsoptimalnumbenchmarkslengthgrtfifty}{4\xspace}
\newcommand{\zdagsizevsoptimalvanillasumlengthgrtfifty}{218\xspace}
\newcommand{\zdagsizevsoptimalsumlengthgrtfifty}{367\xspace}
\newcommand{\zdagsizevsoptimalpercentsmallerthanvanillalengthgrtfifty}{0.0\%\xspace}
\newcommand{\zdagsizevsoptimalpercentsmallerthanhalfvanillalengthgrtfifty}{0.0\%\xspace}
\newcommand{\zdagsizevsoptimalpercentreductionlengthgrtfifty}{-68.3\%\xspace}
\newcommand{\zdagsizevsoptimalpercentasbiglengthgrtfifty}{168.3\%\xspace}
\newcommand{\zdagsizevsoptimalbestpercentasbiglengthgrtfifty}{156.9\%\xspace}
\newcommand{\zdagsizevsoptimalotherpercentasbiglengthgrtfifty}{59.4\%\xspace}

\newcommand{\numherbiebenchmarks}{346\xspace}
\newcommand{\percentgreedyequaloptimal}{81.1\%\xspace}
\newcommand{\numoptimalskipped}{0\xspace}
\newcommand{\numoptimaltried}{3\thinspace760\xspace}
\newcommand{\avesecondsoptimal}{1.24\xspace}
\newcommand{\avesecondsgreedy}{0.02\xspace}
\newcommand{\percentztimeout}{5.0\%\xspace}
\newcommand{\averagemillisegggreedy}{39.57\xspace}
\newcommand{\averageoverheadgreedy}{17.47\xspace}
\newcommand{\averageoverheadproofsenabled}{6.06\xspace}
\newcommand{\averagemillisz}{213.25\xspace}
\newcommand{\averagemillisreduction}{1.47\xspace}
\newcommand{\averagezstartuptimemillis}{17.43\xspace}
\newcommand{\bestdagreductiongreedyvsz}{85.0\%\xspace}
\newcommand{\bestdagreductiongreedyvsvanillaegg}{97.7\%\xspace}
\newcommand{\maxtimeoptimal}{117\thinspace077\xspace}

\newcommand{\numherbiebenchmarksvanilladagsizegrtten}{346\xspace}
\newcommand{\percentgreedyequaloptimalvanilladagsizegrtten}{58.0\%\xspace}
\newcommand{\numoptimalskippedvanilladagsizegrtten}{0\xspace}
\newcommand{\numoptimaltriedvanilladagsizegrtten}{1\thinspace430\xspace}
\newcommand{\avesecondsoptimalvanilladagsizegrtten}{2.03\xspace}
\newcommand{\avesecondsgreedyvanilladagsizegrtten}{0.03\xspace}
\newcommand{\percentztimeoutvanilladagsizegrtten}{9.7\%\xspace}
\newcommand{\averagemillisegggreedyvanilladagsizegrtten}{52.55\xspace}
\newcommand{\averageoverheadgreedyvanilladagsizegrtten}{26.98\xspace}
\newcommand{\averageoverheadproofsenabledvanilladagsizegrtten}{7.83\xspace}
\newcommand{\averagemilliszvanilladagsizegrtten}{545.75\xspace}
\newcommand{\averagemillisreductionvanilladagsizegrtten}{3.85\xspace}
\newcommand{\averagezstartuptimemillisvanilladagsizegrtten}{17.45\xspace}
\newcommand{\bestdagreductiongreedyvszvanilladagsizegrtten}{85.0\%\xspace}
\newcommand{\bestdagreductiongreedyvsvanillaeggvanilladagsizegrtten}{97.7\%\xspace}
\newcommand{\maxtimeoptimalvanilladagsizegrtten}{35\thinspace934\xspace}

\begin{abstract}
  %% The abstract should briefly summarize the contents of the paper in
%% 150--250 words.

Satisfiability Modulo Theory (SMT) solvers
  and equality saturation engines
  must generate proof certificates
  from e-graph-based congruence closure procedures
  to enable verification and conflict clause generation.
Smaller proof certificates
  speed up these activities.
Though the problem of generating proofs of minimal size
  is known to be NP-complete,
existing proof minimization algorithms
  for congruence closure
  generate unnecessarily large proofs
  and introduce asymptotic overhead
  over the core congruence closure procedure.
In this paper, we introduce an $O(n^5)$ time algorithm
  which generates optimal proofs under
  a new relaxed ``proof tree size'' metric
  that directly bounds proof size.
We then relax this approach further
  to a practical $O(n \log(n))$ greedy algorithm
  which generates small proofs
  with no asymptotic overhead.
We implemented our techniques in the
  \egg equality saturation toolkit,
  yielding the first
  certifying equality saturation engine.
We show that our greedy approach in \egg
  quickly generates substantially smaller proofs
  than the state-of-the-art Z3 SMT solver
  on a corpus of \prooflengthnumbenchmarks benchmarks.

\end{abstract}

\IEEEpeerreviewmaketitle

\section{Introduction}
\label{sec:intro}

Congruence closure procedures based on \egraphs~\cite{nelson}
  are a central component of
  equality saturation engines~\cite{eqsat,egg}
  and SMT solvers~\cite{cvc4,z3}.
Sophisticated optimizations like
  deferred congruence~\cite{egg} and
  incremental e-matching~\cite{efficient-e-matching}
  make such tools faster, but also
  make guaranteeing correctness more difficult~\cite{yingyang,fast-proof-checking}.

Engineers sidestep the challenge of
  directly verifying high-performance congruence implementations
  by instead extending procedures to generate
  \textit{proof certificates}~\cite{leo-proofs,fast-proof-checking}.
Proof certificates provide
  the sequence of equalities that
  the congruence procedure used
  to establish that two terms are equivalent.
Clients can safely use results from an
  untrusted procedure by checking its proofs.
For example,
  several proof assistants adopt this strategy
  to provide ``hammer tactics''~\cite{coqhammer}
  which dispatch proof obligations to
  SMT solvers and then reconstruct
  the resulting SMT proofs
  back into the proof assistant's logic,
  thus improving automation without
  trusting solver implementations.

Proof \textit{size} can be especially important
  when extending existing verification tools
  with untrusted solvers.
For example,
  in a case study on
  six Intel-provided Register Transfer Level (RTL) circuit design benchmarks~\cite{hardware-egraph-optimisation},
  an untrusted equality saturation engine
  took under $1$ minute to optimize,
  but the existing verification tool
  took $4.7$ hours to replay and check
  the large proof certificates
  generated by existing techniques~\cite{leo-proofs}.
Unfortunately, finding proofs of minimal size is
  an NP-complete problem~\cite{NP-hardness-small-conflict}.
%%% This is because the existing verification tool
%%%   does not support quantifiers and
%%%   so could not first separately check axioms.
%%% Instead, it had to check whole-circuit equivalence
%%%   after every \textit{axiom instantiation},
%%%   i.e., for each explicit proof step
%%%   where quantified variables have been replaced by concrete terms,
%%%   which made replaying large proofs a performance bottleneck.

In this paper, we explore
  efficient generation of small proof certificates
  for \egraph-based congruence procedures.
We first introduce the problem of \textit{finding minimal size proofs}
  for congruence closure procedures.
We define the space of admissible proofs
  and give an integer linear programming formulation
  for finding a proof with minimal size.
  %by a reduction from the hitting set problem~\cite{garey1979computers}.
Next, we introduce a relaxed metric called
  \textit{proof tree size},
  which directly bounds the size of the proof,
  and develop \treeopt, an $O(n^5)$ time algorithm
  for finding a proof with minimal proof tree size.
Unfortunately, the $O(n^5)$ algorithm
  is still too expensive for
  practical use, since congruence closure procedures
  often consider thousands of equations.
Thus we also developed
  an $O(n \log(n))$ time greedy approach
  using subproof size estimates.
%  an $O(n \log(n))$ time greedy approach
%  using estimates for proofs of congruence closure.
Our algorithm incurs no asymptotic overhead relative to congruence closure
  and finds small proofs in practice.

We evaluate our approach by
  implementing both proof generation
  and greedy proof minimization
  in the state-of-the-art
  \egg equality saturation toolkit~\cite{egg},
  yielding the first
  certifying equality saturation engine.
We compare our greedy algorithm against the state-of-the-art SMT solver Z3,
  which performs proof reduction (see~\Cref{sec:background})
  to find smaller proofs.
Where we can run Z3 (Z3 times out in \percentztimeout of cases),
  our proofs are only \zdagsizepercentasbig as big as Z3's on average
  (\zdagsizebestpercentasbiglengthgrtten in the best case).
Our proofs are also only \optimaldagsizevsgreedyotherpercentasbig
  as big as \treeopt's on average, compared to \zdagsizevsoptimalpercentasbig for Z3.
Using our greedy proof minimizer,
  we were able to reduce proof replaying time
  in the Intel-provided RTL verification case study
  from $4.7$ hours down to $2.3$ hours.

In this paper, we first define the problem of finding the minimal proof and provide an ILP formulation (\Cref{sec:explanation-length}).
We then introduce the proof tree size metric and
an optimal $O(n^5)$ time algorithm for
finding proofs of minimal tree size (\Cref{sec:optimal}).
Finally, we demonstrate a practical greedy algorithm for
finding proofs of small tree size
with no asymptotic overhead (\Cref{sec:greedy}).

\section{Background and Related Work}
\label{sec:background}

%\subsection{Congruence closure and \egraphs}

Congruence is the property that $a = b$ implies $f(a) = f(b)$.
Congruence closure refers to building a model
  of a set of equalities that satisfies congruence;
  these models can be used
  for determining whether other equalities are true
  (as is common in SMT solvers)
  or for finding new equivalent forms of an expression
  (as is common in equality saturation engines).
For example,
  consider the equalities $a + 0 = a$
  and $2 + 2 = 4$;
  a model of these two equalities should permit queries
  like whether $f(a + 0, g(a + 0, 2 + 2))$ has a simpler form
  or whether it is equal to $f(a, g(a, 4))$.

\begin{figure}
  \begin{center}
  \includegraphics[width=0.6\linewidth]{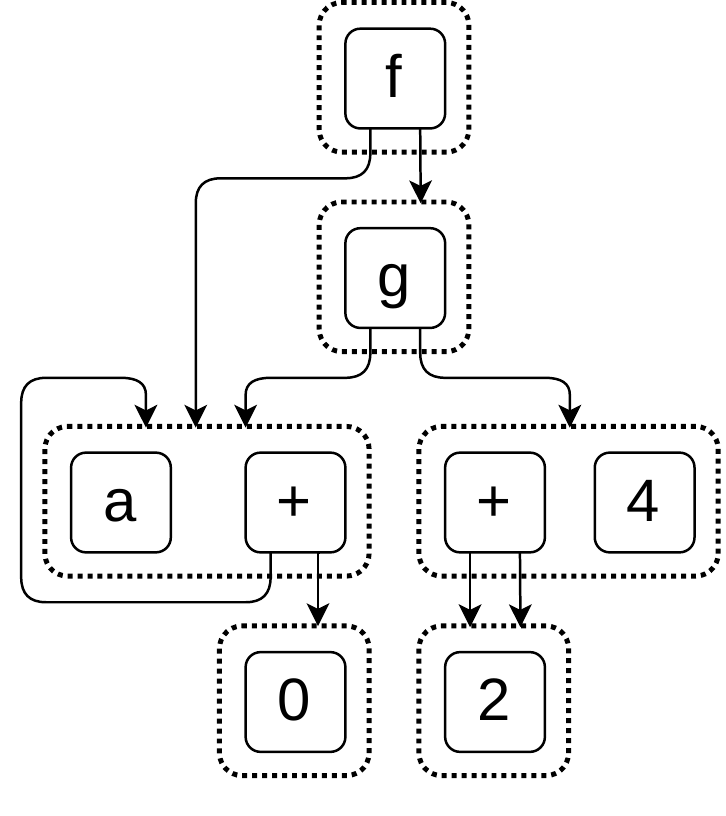}
  \end{center}
  \caption{
    A e-graph model of the equalities
      $a + 0 = a$ and $2 + 2 = 4$
      and the expression $f(a + 0, g(a + 0, 2 + 2))$.
    Note that the top \eclass
      contains both the expression
      $f(a + 0, g(a + 0, 2 + 2))$
      and the expression
      $f(a, g(a, 4))$,
      which proves
      that these two expressions are equal
      modulo the equalities.
  }
  \label{fig:egraph}
\end{figure}

A congruence closure model is typically represented as an \egraph,
  which is a collection of \enodes and \eclasses.%
\footnote{
  Depending on the author, the ``e'' in ``\egraph'' can stand for
    ``expression'', ``equivalence'', or ``equality''.
}
Each \enode represents a single function being applied and
  an \eclass for each argument;
  each \eclass, meanwhile,
  is a set of equivalent \enodes.
Any expression can be inserted into the \egraph
  by converting it recursively into \enodes,
  while equalities can be added into the \egraph
  by merging the \eclasses
  for the equality's left and right hand side.
For example,
  given the equalities $a + 0 = a$ and $2 + 2 = 4$,
  one can determine whether $f(a + 0, g(a + 0, 2 + 2)) = f(a, g(a, 4))$
  by inserting these two expression into an \egraph
  and then adding the two equalities.
The resulting \egraph is shown in \Cref{fig:egraph}.
The two expressions end up in the same \eclass,
  so they have been proven to be equal.

Congruence procedures must handle queries quickly,
  with tens or hundreds of thousands of equalities.
The large number of equalities
  means that \egraphs can contain hundreds of thousands
  or even millions of \enodes,
  with the resulting \egraph taking significant time to construct.
A substantial literature~\cite{egg,efficient-e-matching,relational-ematching}
  describes numerous optimizations to \egraphs.
Past work shows that an \egraph for $n$ equalities can be constructed
  in $O(n \log n)$ time~\cite{DST}.

\mypara{Congruence Proofs}
Proof certificates for \egraphs
  allow checking that two terms are equal
  without reconstructing the \egraph.
Instead, for an equality $E_1 = E_2$ witnessed by the \egraph,
  a proof certificate is a list of given equalities
  that can be applied in order, one after another,
  as rewrite rules to transform $E_1$ into $E_2$.
Some of these equalities are applied
  at the root of the expression being rewritten,
  while others apply to subexpressions (via congruence).
In our running example,
  we can prove $f(a + 0, g(a + 0, 2 + 2)) = f(a, g(a, 4))$
  as follows:

\[ \begin{array}{l}
f(a + 0, g(a + 0, 2 + 2)) \\[4pt]
\;\;\;\;
\;\;
  \xrightarrow{a + 0 = a}
    f(a, g(a + 0, 2 + 2)) \\[4pt]
\;\;\;\;
\;\;\;\;
\;\;
  \xrightarrow{2 + 2 = 4}
    f(a, g(a + 0, 4)) \\[4pt]
\;\;\;\;
\;\;\;\;
\;\;\;\;
\;\;
  \xrightarrow{a + 0 = a}
    f(a, g(a, 4)) \\[4pt]
\end{array} \]

%\[
%f(a + 0, g(a + 0, 2 + 2))
%\xrightarrow{a + 0 = 0}
%f(a, g(a + 0, 2 + 2))
%\xrightarrow{2 + 2 = 4}
%f(a, g(a + 0, 4))
%\xrightarrow{a + 0 = 0}
%f(a, g(a, 4))
%\]

\noindent
Note that some equalities may be reused,
  as in this example.

Over time,
  proof certificates have grown increasingly important.
In SMT solvers,
  proof certificates correspond to conflict clauses
  and enable \textit{non-chronological backtracking},
  a key component of modern SMT solvers \cite{smt-backtracking}.
In proof automation,
  proof certificates bridge foundational logics
  and unverified automated theorem provers,
  as in the ``hammer'' style of proof tactics~\cite{coqhammer}.
In equality saturation engines,
  replaying proof certifications
  enables the combination of slow verification procedures
  with fast equality saturation engines.

To produce proofs certificates, \egraph implementations
  maintain a spanning tree for each \eclass,
  with each edge of the tree justifying the equality
  of the two \enodes it connects~\cite{pp-congr}.
This justification is either
  one of the (quantifier-free) equalities provided as input
  or a congruence edge that refers to
  other connected nodes in the tree.
This spanning tree is maintained alongside the
  union-find structure used for efficiently merging \eclasses,
  so there is no algorithmic overhead to maintaining it.
Producing a proof for the equality of two \enodes in the same \eclass
  is then a simple recursive procedure which
  traverses the path between two \enodes,
  recursively finding subproofs for each congruence edge.
In a spanning tree,
  there is a unique path between any two \enodes,
  so this recursive algorithm is quite fast,
  taking $O(n \log n)$ time for $n$ equalities.

\mypara{Shrinking Congruence Proofs}
Most uses of proof certificates,
  including generating conflict clauses
  and replaying and checking proofs,
  take longer as more
  unique equalities are used in the proof certificate.
The standard approach to finding smaller proof certificates,
  implemented in SMT solvers such as Z3~\cite{z3},
  is based on the observation~\cite{pp-congr}
  that proof certificates can contain redundant equations;
  for example, if the given equalities include
  $a = b$, $a = c$, and $b = c$,
  a proof certificate may include all three.
By attempting to re-prove the same equation
  while excluding one of the equalities,
  a proof certificate can thereby be shrunk.
If the initial proof certificate has length $k$,
  this proof reduction procedure takes $O(k^2 \log k)$
  (as checking the validity of each new proof takes
  $O(k \log k)$ time using an \egraph).

This state of the art algorithm is limited in two ways.
First, when $k \in o(\sqrt{n}) $,
  it introduces an asymptotic slowdown
  over the rest of the congruence closure algorithm,
  which can answer queries and generate proofs
  in $O(n \log n)$ time (where $n$ is the number of equalities).
Second and more importantly,
  proof reduction is ultimately limited by the choice of the proof to reduce.
Since proof reduction is too slow to consider the entire \egraph,
  a valid initial proof is generated before applying proof reduction,
  discarding many (potentially useful) equalities right away.
This means that, while it results in shorter proof certificates,
  those proof certificates are still longer than optimal.
This paper addresses both concerns.

\section{Optimal DAG Size}
\label{sec:explanation-length}

%% The ``right'' metric for proof size
%%   depends on how the proof will be checked.
%%
%%
%% In many scenarios,
%%   an equality saturation engine or SMT solver
%%   applies e-matching to instantiate
%%   grounded equalities from a set of
%%   quantified axioms $A$.
%% These grounded equalities are then
%%   used for computing congruence closure.
%% If the client trusts all the
%%   quantified axioms in $A$,
%%

Because proof certificates often
  contain repeated subproofs,
  we propose a measure for a proof's size in terms of
  the number of \textit{unique} equalities
  it uses.
We call this measure \textit{DAG size} because equalities may be reused in
  the proof.
DAG size is also the same as the size of a conflict set in the context
  of SMT solvers.
The problem of finding a proof of minimal DAG size is also NP-complete~\cite{NP-hardness-small-conflict}.
% In this section, we formalize the problem and give a linear programming formulation for it.
This section formalizes
  a \textit{DAG size} measure of proof length
  which accounts for subproof reuse,
  and gives an ILP formulation for finding the proof of optimal DAG size.

%% We are interested in minimizing proof length
%%   for translation validation or conflict clause production.
%% \of{we haven't said translation validation anywhere yet- maybe proof checking or replaying?}
%% In this case, shared subproofs can be checked only once,
%%   and we wish to minimize the total number
%%   of unique instantiations of our axioms.
%% This section formalizes and defines
%%   this \textit{DAG size} measure of proof length,
%%   and shows that minimizing it is NP-hard.

\subsection {C-graphs}

Traditionally, each equivalence class in an e-graph is represented by a spanning tree.
Each edge in the spanning tree is either a single equality between two terms or equality via congruence.
Any additional equalities between nodes already connected are discarded, since there is already a way to prove the two terms are equal.
However, these equalities may enable a significantly smaller proof.
For example, an e-graph can be constructed from the equalities $a = b$, $b = c$, and $a = c$.
The e-graph constructs a spanning tree with edges $a = b$ and $b = c$, discarding $a = c$.
Now the e-graph will admit a proof between $a$ and $c$ that has a size of $2$.

Since these additional equalities can be used to produce shorter proofs, our algorithm requires storing them.
We call the resulting structure a \cgraph, which maintains a graph, not a spanning tree, for each equivalence class.
Storing these additional edges merely requires recording information on every e-graph merge operation, so can be done without changing the complexity of the congruence closure algorithm.
The \cgraph can be substituted directly for an \egraph without changing the complexity 
of the congruence closure algorithm.
In practice, a \cgraph uses the same representation and algorithms as an \egraph, but additionally has an adjacency list for each node storing this graph of equalities.
In the context of producing proofs, we define a simple version of a \cgraph below:

\begin{definition}
  A \cgraph is an undirected graph $G = (V, E)$,
    where nodes $V$ represent expressions
    and edges $E$ represent equalities,
    along with a \textit{justification} $j(e)$ for edge $e$.
  A justification is either an equality $v_1 = v_2$ between the vertices
    or a congruence subproof $c_1 = c_2$, where $c_i$ is a child of $v_i$.
\end{definition}

For convenience, we write $C$
  for the set of congruence edges in $E$.
An edge justified by an equality
  connects the left and right-hand sides of the equality directly,
  while an edge justified by a congruence $c_1 = c_2$
  connects terms which are equal by congruence over $c_1$ and $c_2$
  (e.g. $f(c_1)$ and $f(c_2)$).
If two terms are equal due to the congruence of multiple children,
  the \cgraph contains one congruence edge per argument (one per child).
This keeps the encoding simple, as each congruence edge corresponds to one proof of congruence.
All functions have a bounded arity,
  so this transformation does not affect complexity results.

\begin{figure}
  \includegraphics[trim=40 650 60 60,clip,width=\linewidth]{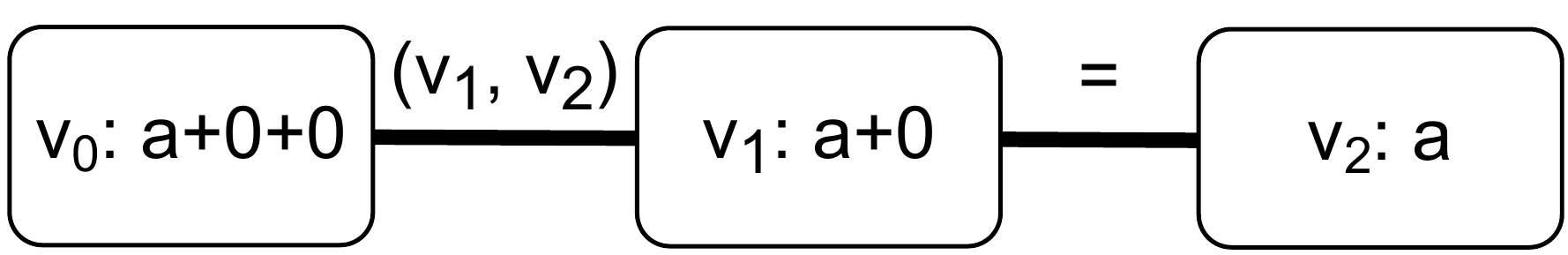}
  \caption{
    A \cgraph proof that
      $a+0+0 = a$.
%    A proof of equality
%      between $a+0+0$ and $a$
%      represented as a \cgraph.
    There is one congruence edge $(v_0, v_1)$
      with $j((v_0, v_0)) = (v_1, v_2)$.
    Since $v_0$ and $v_2$ are e-connected, the proof holds.
  }
  \label{fig:graph-congruence-example}
\end{figure}

For a \cgraph to be a valid proof,
  all congruence edges must refer to e-connected nodes:

\begin{definition}
  \label{def:valid-congruence}
  A congruence edge $e \in E$ with $j(e) = (c_1 = c_2)$ is \textbf{valid}
  if the congruent children $c_1$ and $c_2$ are \textbf{e-connected}
  in the reduced \cgraph $(G', j)$, where $G' = (V, E \setminus \{e\})$.
  All non-congruence edges are valid.
\end{definition}

\begin{definition}
  \label{def:e-connectivity}
  Two vertices $v_s$ and $v_t$ are \textbf{e-connected} in a \cgraph $(G,j)$
  if there is a path between them consisting of \textbf{valid} edges in $E$.
\end{definition}

A \cgraph then proves $s = t$
  if the corresponding vertices $v_s$ and $v_t$ are e-connected.
The particular path showing that $v_s$ and $v_t$ are e-connected, along with proofs
  for each congruence edge along the path, represents a particular proof.
The definition of e-connectedness and edge validity are mutally recursive;
  the base case occurs when two vertices are connected by a set of
  non-congruence edges.

The \cgraph structure allows for
  a simple definition of the DAG size metric:

\begin{definition}
  The \textbf{DAG size} of a \cgraph $(G, j)$ is $| E \setminus C |$,
    the number of non-congruence edges it contains.
\end{definition}

Each non-congruence edge $e \in E \setminus C$ could also be
  assigned a positive, real-numbered weight $w(e)$, giving a weighted DAG size:
  $\sum_{e \in E \setminus C} w(e)$.
Applications could leverage these weights in order to sample proofs
  that minimize an alternative objective function, such as
  the run-time of verifying the steps of the proof.
The algorithms in this paper easily support
  weighted DAG size, but we will use the simpler
  definition of DAG size with each non-congruence edge assigned a weight of 1.

\subsection{Minimal DAG Size}

\begin{figure*}
  \[
  \begin{array}{lll}
  \textsc{Edges} &
  S[i, j] \le (i, j) \in E \setminus C &
  S[i, j] = S[j, i] \\[12pt]
  \textsc{Congruence} &
  M[i, j, l, r] \le (i, j) \in E \land j((i, j)) = (l = r) &
  M[i, j, l, r] = M[j, i, r, l] \\[12pt]
  \textsc{Paths} &
  P[i, i, j] = 0 &
  P[i, k, j] \le V[i, j] \\[5pt]
  &
  C[i, j] = \sum_k P[i, k, j] &
  P[i, k, j] \le C[k, j] \\[12pt]
  \textsc{Validity} &
  V[i, j] \le S[i, j] + \sum_{l, r} M[i, j, l, r] \\[12pt]
  \textsc{No Cycles} &
  0 \le D[i, j] \le \ell &
  D[i, j] \ge 1\text{ if }i \ne j \\[5pt]
  &
  (1 - P[i, k, j]) \ell + (D[i, j] - D[k, j]) \ge D[i, k] \\[5pt]
  &
  (1 - M[i, j, l, r]) \ell + D[i, j] \ge D[l, r] \\[12pt]
  \textsc{Goal} &
  C[v_s, v_t] = 1 &
  \min \sum_{i, j} S[i, j] \\
  \end{array}
  \]
  \caption{
    An integer linear programming formulation
      of the minimum DAG size problem.
    Variables $S$, $M$, $V$, and $P$ are sets of boolean variables,
      while $D$ is integer-valued.
    Variables are indexed by $i$, $j$, and $k$, which represent
      nodes in the \cgraph.
    Decision variables $S$ and $M$ define
      which non-congruence and congruence edges of $E$ are selected respectively.
    $\ell = |C|^{|C| + 1} |E|$ bounds the maximum length
      of a valid non-cyclic path.
  }
  \label{fig:ilp}
  \end{figure*}
The key to finding shorter proofs is to keep track of
  a \cgraph of possible proofs during congruence closure,
  from which a short proof can eventually be extracted.
Traditional congruence closure algorithms
  store only one proof of equality between any two terms
  (they generate \cgraphs shaped like forests)
  because they discard any equalities they discover
  between already-equal terms.
Instead, we will store these redundant edges,
  producing a \cgraph shaped like a full graph,
  and will then later search this \cgraph
  for a sub-\cgraph of minimal size.
We will also discover any extra opportunities for proofs
  of congruence between terms, adding these
  to the \cgraph as congruence edges.

\begin{definition}
  Consider a \cgraph $(G, j)$, all of whose edges are valid.
  We write $(G', j) \subseteq (G, j)$
  when $G' \subseteq G$ and all edges in $(G', j)$ are valid.
\end{definition}

The goal is then to find the sub-\cgraph of minimal size
  in which two terms $s$ and $t$ remain e-connected.

\begin{definition}[The Minimum DAG size Problem]
  Given a \cgraph $(G, j)$ and two e-connected terms $s$ and $t$,
  find a $(G', j) \subseteq (G, j)$ in which $s$ and $t$ remain e-connected
  with minimal DAG size.
\end{definition}

Note that a sub-\cgraph is defined by
  which edges in $G$ it keeps;
  this allows us to phrase the minimum DAG size problem
  as an integer linear programming problem
  with one decision variable per edge in $E$.
The full linear programming problem
  is given in \autoref{fig:ilp}.
It defines selected edges via $S$ and $M$,
  paths $P$ and e-connectedness $C$
  (via edge validity $V$),
  and breaks cycles using distance measure $D$;
  it is similar to the standard formulation
  of graph connectedness as an ILP problem,
  except with extra constraints
  for the validity of congruence edges.
These constraints require the selected edges $S$ and $M$
  to form a sub-\cgraph of $(G, j)$
  with all edges valid.
Finally, $v_s$ and $v_t$ are asserted to be e-connected
  to ensure that the sub-\cgraph proves $s = t$
  and then DAG size is minimized.
While this ILP formulation is solvable
  by industry-standard ILP solvers
  for very small instances,
  it is NP-complete in general~\cite{NP-hardness-small-conflict}.

\section{Optimal Tree Size}
\label{sec:optimal}

What makes the minimal DAG size problem NP-complete
  is the fact that the e-connectedness of multiple congruence edges
  can rely on the same edges.
This sharing means that the cost of using a congruence edge
  depends on equalities other congruence edges rely on---%
  global information about the sub-\cgraph of the solution as a whole.
Instead of finding the optimal solution,
  we optimize for a different metric
  to achieve a practical algorithm
  for proof length minimization.
The distance metric $D[i, j]$ in the ILP formulation,
  which we call the \textit{tree size} of a \cgraph,
  is an effective metric for this purpose.

The tree-size of a \cgraph
  is computed by summing the length of the proof,
  without sharing.
Specifically, given a \cgraph $(G, j)$ that proves $s = t$,
  its tree size is the tree size of the path
  from $v_s$ to $v_t$:
\begin{definition}
  Consider a path $P$ that e-connects $v_i$ to $v_j$
    in a \cgraph.
  The tree size of $P$
    is the number of non-congruence edges in $P$
    plus, for each congruence edge justified by $(v_l = v_r)$,
    the tree size of the path from $v_l$ to $v_r$.
\end{definition}

If a \cgraph has minimal DAG size,
  its DAG size is the number of non-congruence edges in the graph.
Its tree size, meanwhile,
  may count each more than once,
  so presents an upper bound on the DAG size.%
\footnote{
  We chose the name ``DAG size'' and ``tree size''
  because the relationship between these two metrics
  is similar to the relationship between a DAG and a tree
  containing the same parent-child relationships.
}
We can thereby hope that the \cgraph of minimal tree size
  will also have a small DAG size.

\begin{definition}[The Minimum Tree Size Problem]
  Given a \cgraph $(G, j)$ that proves $s = t$,
  find the $(G', j) \subseteq (G, j)$ that proves $s = t$
  and has minimal tree size.
\end{definition}

\subsection{Minimum \Proof Tree Size Algorithm}

Unlike DAG size,
  tree size does not have the problem of shared edges.
Finding a proof of optimal tree size
  thus does not require global reasoning about the surrounding context:
  using the same edges with another part of the proof
  does not reduce the tree size.
As a result, it is possible to solve the minimum tree size problem
  in polynomial time.

Finding a proof of optimal tree size is not a simple graph search.
The key problem is that congruence edges
  may contain other congruence edges in their subproofs,
  and the tree size of those subproofs is initially unknown.
Moreover, often a congruence edge $(v_1, v_2)$
  can be proven in terms of another congruence edge $(v_3, v_4)$
  and vice versa.
Our algorithm tackles this problem by computing the size of proofs
  of congruence bottom up,
  in multiple passes.
At the $i$-th pass,
  it constructs proofs of equalities between vertices
  where congruence subproofs only go $i$ layers deep.
These proofs form an upper bound on the optimal tree size,
  decreasing in size until the optimal proof is found.
When the algorithm reaches a fixed point,
  the \eproof of optimal tree size is discovered.
The algorithm for finding the size of the optimal proof
is given in \autoref{fig:flattened-optimal-code}.
With more bookkeeping, it can be easily extended
to yield the specific proof the optimal size corresponds to.

\begin{figure}
  \begin{minipage}[t]{1\linewidth}
    \begin{lstlisting}[
      gobble=4,
      numbers=left,
      basicstyle=\scriptsize\ttfamily,
      mathescape=true ]
    def optimal_tree_size(start, end):
      for i in G.vertices:
        dist[(i, i)] = 0

      for ($\ell$, $\rrr$) in E \ C:
        dist[$\ell$, $\rrr$] = 1

      for i in range(|C|):
        for ($\ell$, $\rrr$) in C:
          dist[$\ell$, $\rrr$] = shortest_path($\ell$, $\rrr$, dist)
      return shortest_path(start, end, weights=dist)
    \end{lstlisting}
  \end{minipage}
  \caption{
      Pseudocode for the optimal proof tree size algorithm.
      The algorithm keeps a dictionary $dist[a, b]$,
      the length of the shortest tree size from $a$ to $b$ found so far.
    }
  \label{fig:flattened-optimal-code}
\end{figure}

In each pass, this algorithm computes the shortest path for each
  proof of congruence.
Non-congruence edges have a weight of 1,
  and congruence edges are initialized to have infinite weight.
A fixed point is guaranteed after $|C|$ iterations,
  because each subproof for a congruence edge $e$
  cannot use the same edge $e$ again
  (else its tree size would increase).
The overall running time of the algorithm
  is bounded by $O(|C|^2 |E|)$,
  with $|C|^2$ being the number of calls to the shortest path algorithm
  and $|E|$ being the complexity of finding a shortest path given the weights.
Since there may be $n^2$ congruence edges for $n$ nodes in the graph,
  the overall running time is also bounded by $O(n^5)$.
However, in practice the number of congruence edges is some constant
  multiple of $n$, and in this case the running time is $O(n^3)$.

\section{Greedy Optimization of \Proof Tree Size}
\label{sec:greedy}

The optimal algorithm of \Cref{sec:optimal}
  finds the proof with minimal tree size,
  but it does so at an unacceptable cost:
  its running time dominates
  the $O(n \log n)$ running time of congruence closure itself~\cite{nelson}.
In the context of c-graphs, $n = |E \setminus C|$, the set of input
  equalities to congruence closure.
This section thus proposes
  a greedy algorithm for \eproof tree size,
  which reduces tree size and DAG size significantly in practice,
  though it is not optimal with respect to either metric.

\subsection{Greedy Optimization}

The key insight behind the greedy algorithm
  is that the multiple passes of the optimal algorithm
  are only necessary to compute the minimal cost of congruence edges.
If the tree size for each congruence edge were known,
  the proof with optimal tree size could be found
  by a simple shortest path algorithm.
The greedy algorithm is a simple
  breadth-first search shortest path algorithm
  that takes estimated costs for congruence edges as an input.
The closer the estimates are
  to the proof of optimal tree size,
  the better the results of the greedy algorithm.

Defer for now the challenge
  of estimating the tree size for each congruence edge,
  and focus on the greedy algorithm itself.
The algorithm is simple:
  use a breadth-first search to
  choose a path from the start vertex $s$
  to the end vertex $t$ of minimal length,
  using the estimates for each congruence edge.
However, those estimates may not be optimal,
  so the algorithm then recurses for each congruence edge.
Note the difference between the optimal algorithm
  (which first optimizes congruence edges)
  and the greedy algorithm
  (which first finds a shortest path).
If the recursion were performed
  until all congruences are optimized,
  this algorithm would take time $O(|C| (n + |C|))$,
  which is still too high compared to the $O(n \log(n))$ runtime of congruence closure.
Instead, only $T$ expansions of congruence edges are permitted;
  in practice, we choose $T=10$,
  which seems to work well.
After $T$ expansions, there may be sub-proofs which have not been generated.
In this case, the algorithm defaults to a generic
  proof production algorithm for the remaining sub-proofs~\cite{pp-congr}.
\autoref{fig:greedy-optimization} lists the greedy algorithm.

\begin{figure}
  \begin{minipage}{1\linewidth}
    \begin{lstlisting}[
      gobble=4,
      numbers=left,
      basicstyle=\scriptsize\ttfamily,
      escapechar=|,
      mathescape=true ]
    def greedy(start, end, pf_size_estimates):
      todo = Queue((start, end))
      fuel = |$T$|

      while len(todo) > 0:
        (start, end) = todo.pop()
        path = shortest_path(start, end, pf_size_estimates)
        for edge in path:
          match edge:
            congruence($\ell$, $\rrr$) ->
              if fuel > 0:
                todo.push($\ell$, $\rrr$)
                fuel = fuel - 1
              else:
                add_to_proof(unoptimized_proof($\ell$, $\rrr$))
            axiom(a) ->
              add_to_proof(a)
    \end{lstlisting}
  \end{minipage}
  \caption{
      Pseudocode for the greedy optimization of proof tree size.
      The algorithm either recurs for congruence edges if fuel allows, or it uses the estimates for each congruence edge.
      Unlike \treeopt, the algorithm is top-down and terminates after $T$ steps.
    }
  \label{fig:greedy-optimization}
\end{figure}

\subsection{Estimating Tree Sizes}

The main challenge to instantiating the greedy algorithm
  is generating size estimates for congruence edges.
However, there is a simple way to do so:
  reduce the \cgraph to a forest $(G_t, j)$
  with one tree per connected component,
  in such a way that all edges remain valid.
Luckily, the traditional congruence closure proof production algorithm
  generates such reduced \cgraphs
  by omitting any unions which connect already-equal terms.
Now, the tree size of a proof of congruence can be estimated by directly
  calculating the tree size of a proof in the reduced instance.
In such a reduced \cgraph,
  there is only one possible path between any two nodes,
  so the proof is unique.

\begin{figure}
  \begin{center}
  \includegraphics[width=0.6\linewidth]{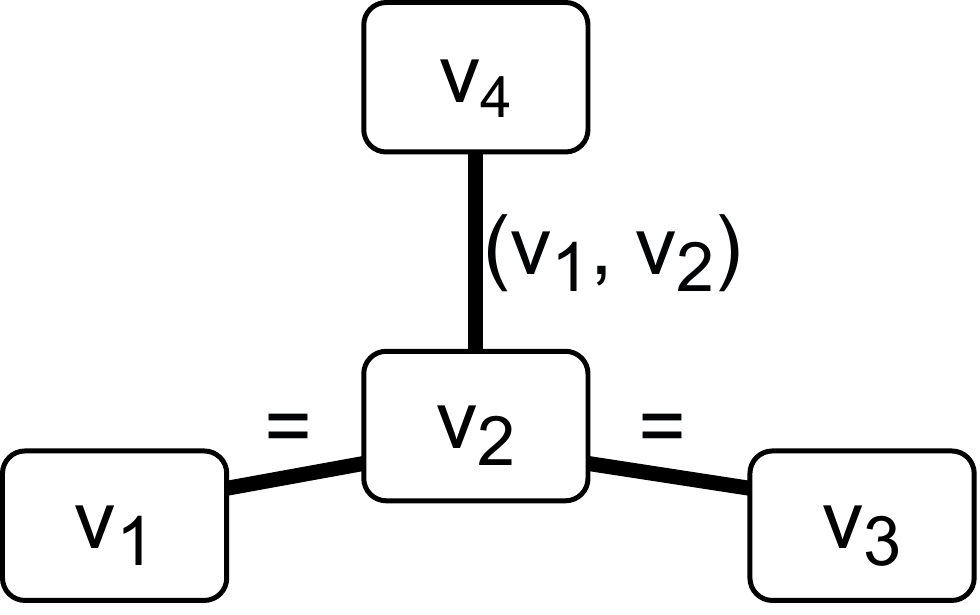}
  \end{center}
  \caption{
    An example reduced c-graph with a single congruence edge.
    The root of the tree is the vertex labeled $v_4$ at the top,
    and there is a single congruence edge $(v_1, v_2)$ in the
    spanning tree.
    The proof of congruence between vertices $1$ and $2$ has a tree size of two
    because the proof between the congruent children involves
    two equalities.
  }
  \label{fig:congruence-cycle-example}
\end{figure}

Computing the tree sizes of all proofs in the reduced \cgraph
  requires some care to stay within the necessary asymptotic bounds.
First, each tree in $(G_t, j)$ is arbitrarily rooted.
Given a vertex $a$,
  let \texttt{size[$a$]} be the size of the proof between $a$
  and the root of its tree.
Then the tree size of the proof between any two vertices $a$ and $b$
  can be calculated
  \[\texttt{size[$a$] + size[$b$] - 2 * size[lca($a$, $b$)]},\]
  where \texttt{lca} computes
  the least common ancestor of $a$ and $b$ in the tree.
The \texttt{lca} function can be pre-computed for all relevant proofs in $O(n)$ time
  using Tarjan's off-line algorithm \cite{tarjan-off-line-lca}.

\begin{figure}
  \begin{minipage}[t]{1\linewidth}
    \begin{lstlisting}[gobble=4, numbers=left, basicstyle=\scriptsize\ttfamily, escapechar=|]

    def path_compress(vertex):
      if parent[vertex] != vertex:
        path_compress(parent[vertex])
        parent[vertex] = parent[parent[vertex]]
        size[vertex] = size[vertex] + size[parent[vertex]]

    def traverse_to_ancestor(v, ancestor):
      while parent[vertex] != ancestor:
        edge = parent_edge(parent[vertex], G)
        parent[edge.start] = edge.end
        if is_congruence(edge):
          traverse(j(edge).start, j(edge).end)
        estimate_size(edge)
      path_compress(vertex)

    def traverse(start, end):
      path_compress(start)
      path_compress(end)
      ancestor = argmin(
        (lca(start, end), parent[start], parent[end]),
        distance_to_root)
      path_compress(ancestor)

      # Ensure that start, end, and their lca share a parent
      traverse_to_ancestor(start, ancestor)
      traverse_to_ancestor(end, ancestor)
      estimate_tree_size(start, end)

    def estimate_tree_size(start, end):
      tree_size[(start, end)] = size[start] + size[end]
                                - 2*size[lca(start, end)]

    def estimate_size(edge):
      match edge:
        congruence(left, right) ->
          size[edge.start] = tree_size[(left, right)]
        axiom(a) ->
          size[edge.start] = 1

    for i in G.vertices:
      parent[i] = i
      size[i] = 0

    for (start, end) in congruence_edges(G):
      traverse(start, end)

    \end{lstlisting}
  \end{minipage}
  \caption{
      Pseudocode for computing tree sizes of all congruence proofs given
      $(G_t, j)$.
      The algorithm efficiently computes these tree sizes by storing a union-find datastructure
        that keeps track of \texttt{size}, the size of the proof between a node and it's parent.
      Computing the size of a proof involves traversing the proof,
        updating the union-find whenever the size of a sub-proof is discovered.
      The pseudocode uses the function \texttt{distance\_to\_root} to denote the number
      of edges from $v$ to the root of its tree.
      It also makes use of \texttt{lca}, a function that returns the lowest common
      ancestor of two vertices.
    }
  \label{fig:flattened-greedy-estimation}
\end{figure}

\autoref{fig:flattened-greedy-estimation}
  shows the pseudocode for calculating proof tree sizes given $(G_t, j)$.
To avoid an infinite loop in proof length calculation,
  the algorithm builds each tree in $(G_t, j)$ incrementally
  using a union-find structure (using the \texttt{parent} array).
Consider the example in \autoref{fig:congruence-cycle-example}, in which the path
  to the root node $v_4$ contains a congruence edge.
The tree size of the proof between nodes $v_2$ and $v_4$, written \texttt{tree\_size($v_2$, $v_4$)},
  involves calculating the size of the congruence
  proof \texttt{tree\_size($v_1$, $v_3$)}.
So \texttt{tree\_size($v_2$, $v_4$)} cannot be computed using $v_4$ as the root of the tree,
  since the path to the root involves the congruence edge.
Instead, the algorithm uses least common ancestor $v_2$
  to compute \texttt{tree\_size($v_1$, $v_3$)}.
Because the proof is e-connected, any congruence edges on the path to the least
  common ancestor can be computed recursively without diverging.

Each congruence edge results in at most one recursive call to
  \texttt{traverse}, while non-congruence edges are added to the union-find
  data structure directly.
Ultimately, each edge in the \cgraph
  contributes at most five union-find operations:
  three \texttt{find} operations at the start of \texttt{tree\_size},
  one \texttt{union} operation to add it to the union-find data structure,
  and one more \texttt{find} in \texttt{traverse\_to\_ancestor}.
A sequence of $m$ operations
  on a union-find data structure with $h$ nodes
  can be executed in $O(m \log(h))$ time~\cite{unionfind}.
This means the overall cost of estimating sizes for congruence edges
  is $O(n \log(n))$ since $n$ bounds both $m$ and $h$ (recall $n = |E \setminus C|$).
Adding on $O(n + |C|)$ cost for the greedy algorithm itself
  yields an overall runtime of $O(n \log(n) + n + |C|) = O(n \log(n) + |C|)$.
Limiting the number of congruence edges $C$
  to a multiple of $n$
  results in a $O(n \log(n))$ runtime, introducing no asymptotic
  overhead compared to congruence closure alone.
\footnote{In practice, $|C|$ is typically
  a small constant factor larger than $n$.
We use a constant factor of $10n$ as a reasonable limit on the number of congruence edges.}

\section{Evaluation}
\label{sec:eval}

\begin{figure}
  \includegraphics[width=\linewidth]{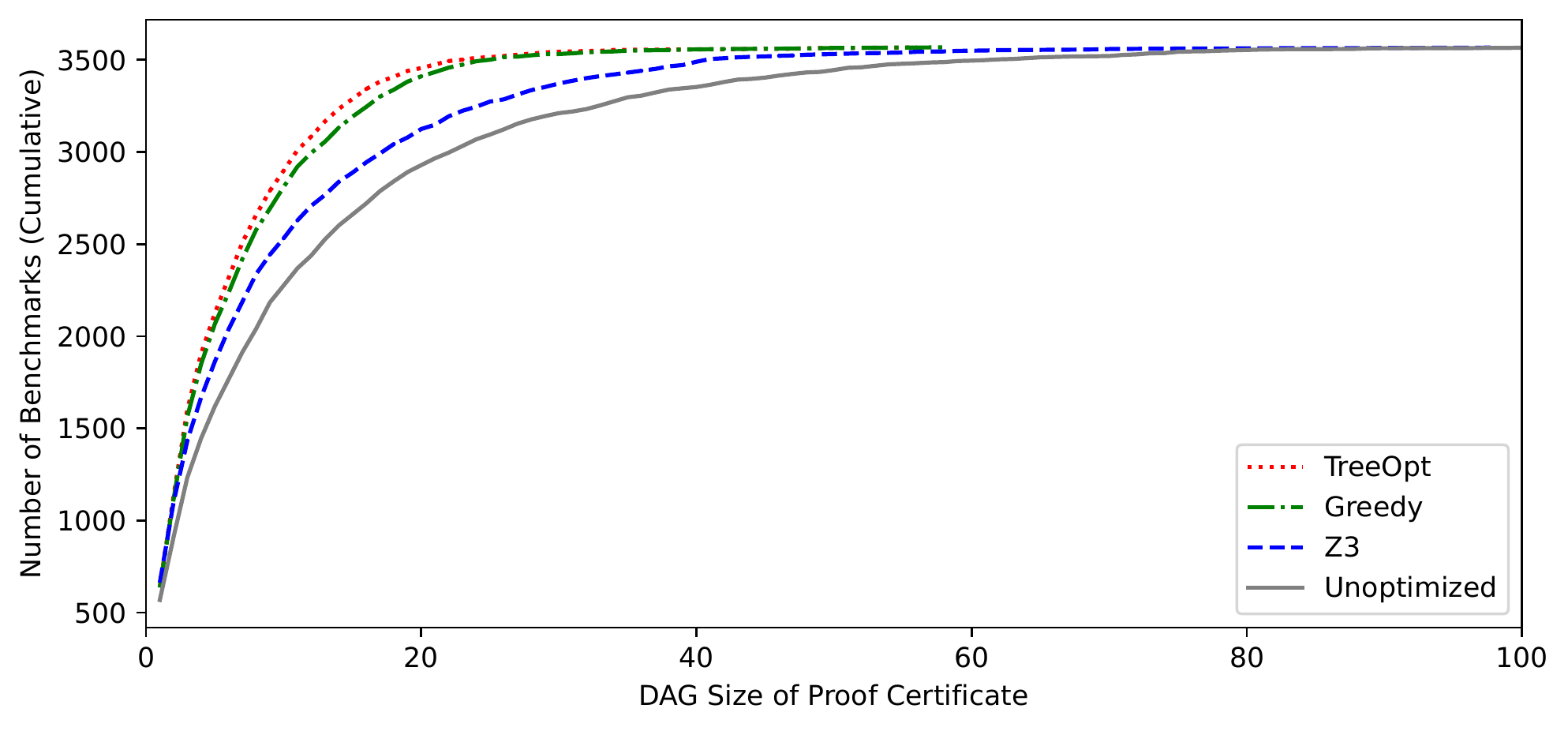}
\caption{
  This CDF compares the
    unoptimized (gray solid),
    Z3 (blue dashed),
    greedy (green dash-dotted),
    and \treeopt (red dotted)
    proof generation algorithms
    on the same \zdagsizenumbenchmarks benchmarks
    where Z3 does not time out.
  Each line shows the number of benchmarks
    whose proofs are at most
    the size indicated on the horizontal axis.
  Our greedy approach (green)
    closely tracks the size of \treeopt's (red) proof certificates,
    showing that its certificates are difficult to shrink further.
  Five outliers with an unoptimized DAG size of more than 100 are omitted.
  }
  \label{fig:dag-size-cdf-optimal}
\end{figure}

This section compares
  an implementation of
  our greedy proof generation algorithm
  in the \egg equality saturation toolkit~\cite{egg}
  to Z3's proof generation~\cite{proofs-refutations-z3}.
As described in \Cref{sec:background},
  Z3 applies \textit{proof reduction}
  to the first proof it finds,
  which substantially reduces proof size.
Our greedy approach
  instead attempts to extract a
  minimal proof from the \egraph.
We found that,
  even without a proof reduction post-pass,
  our greedy approach can quickly find
  significantly smaller proofs than Z3
  (\Cref{fig:dag-size-cdf-optimal}).

%% This paper's proof optimization algorithm
%%   allows congruence closure procedures to produce short proofs
%%   for replay and checking.
%% To evaluate its algorithms,
%%   we devised a series of experiments
%%   comparing proof-producing congruence closure
%%   as implemented in the Z3 theorem prover~\cite{proofs-refutations-z3}
%%   to our implementation of this paper's algorithms
%%   in the \egg equality saturation library~\cite{egg}.
%% As discussed in \Cref{sec:background},
%%   Z3 implements \textit{proof reduction}, which
%%   substantially reduces the size of proofs after
%%   they have been found.
%% However, using the greedy algorithm introduced in \Cref{sec:greedy},
%%   the \egg equality saturation library typically produces proofs
%%   significantly shorter than Z3.

\subsection{Comparing \egg to Z3}
\label{subsec:compare}

We use Z3 version~4.8.12
  and egg version~0.7.1 compiled with Rust~1.51.0.
\egg is a state-of-the-art equality saturation library
  that implements the rebuilding algorithm
  for speeding up equality saturation workloads.
It is used by projects like Herbie~\cite{herbie},
  Ruler~\cite{ruler} and Szalinski~\cite{szalinski}.
Z3 is a state-of-the-art automated theorem prover
  and is optimized for theorem proving workloads.
To create a realistic benchmark set,
  we used the Herbie~1.5
  numerical program synthesis tool~\cite{herbie}.
Herbie uses equality saturation for program optimization
  and comes with a standard benchmark suite of programs
  drawn from textbooks, research papers, and open-source software.
We extracted Herbie's set of quantified equalities
  and recorded all inputs and outputs from its equality saturation procedure.
This results in \dagsizenumbenchmarks input/output pairs, of which we focus on the  \zdagsizevsoptimalnumbenchmarks where Z3 did not produce an answer after 2 minutes.

For the Z3 baseline, we converted each input/output pair
  into a satisfiability query
  by asserting each quantified equality
  (with a trigger for the left hand side of the equality)
  and then asserting that the input and output are not equal.
Z3 then attempts to prove the input and output are equal
  using an e-graph and the quantified equalities
  (the theory of uninterpreted functions).
We then computed the DAG size by counting
  the number of calls to its \texttt{quant-inst} command~\cite{isabelle-z3-proofs}
  in its proof scripts.
We ran \egg exactly how it is used by Herbie,
  and then optimized proof length using the greedy algorithm of \Cref{sec:greedy}
  and measured DAG size by counting proof nodes.
Z3 times out after 2 minutes for \percentztimeout of the input/ouput pairs,
  and completes in \averagemillisz milliseconds on average for the remainder.
\egg does not time out,
  and runs for an average of \averagemillisegggreedy milliseconds.
To measure DAG size for the resulting proofs,
  we ran both \egg and Z3 in proof-producing mode
  and examined the resulting proofs.

\Cref{fig:dag-size-cdf-optimal} contains the results:
  the proofs produced by \egg
  are \zdagsizepercentasbig as big as Z3's on average,
  despite Z3's use of a proof reduction algorithm.
Moreover, the effect of proof length optimization
  is greater for longer proofs:
  queries with Z3 DAG size over 10 see an average
  \zdagsizepercentreductionlengthgrtten reduction,
  while queries with Z3 DAG size over 50 see an average
  \zdagsizepercentreductionlengthgrtfifty reduction.

  \begin{table}
    \caption{
      Data comparing \egg to Z3
        using different proof production algorithms:
        \egg with proofs of optimal tree size,
        \egg with greedy optimization,
        \egg with traditional proof reduction (see \autoref{sec:background}),
        Z3,
        and \egg without any optimization.
  %%  We write $O^*$ to omit a factor of $n \log(n)$ implicit in the congruence closure procedure,
  %%  focusing instead on the new factors introduced by the proof production algorithm.
      Note that proof reduction's analysis
        is in terms of $k$,
        the size of the unoptimized proof,
        while $n$ is the size of the entire c-graph instance.
      In practice,
        $k$ is often small relative to $n$.
    }
    \begin{tabular}{l r r c p{4cm} }
Algorithm
 & TreeOpt
 & Ave Time (ms) &
 Complexity \\[2pt]
\hline
TreeOpt & 100.0\% & 1008.60 & $O(n ^ 3)$ \\
Greedy & 105.9\% & 39.33 & $O(n \log(n))$ \\
Egg Reduc. & 138.7\% & 23.01 & $O(n \log(n) + k^2 \log(k))$ \\
Z3 & 147.3\% & 130.69 & $O(n \log(n) + k^2 \log(k))$ \\
Egg & 185.9\% & 22.15 & $O(n \log(n))$ \\
\hline 
 \end{tabular}
    \label{fig:algorithms-table}
  \end{table}

  \begin{table*}
    \caption{RTL design benchmark results. Total runtime includes equality
    saturation and proof production runtimes but excludes any formal verification time.}
    \begin{center}
    \begin{tabular}{l r r r r r r r r r}
      %\hline
      & \multicolumn{3}{c}{Tree Size}
      & \multicolumn{3}{c}{DAG Size}
      & \multicolumn{3}{c}{Runtime (sec)} \\[2pt]
      Benchmark {\color{white}.}
        & {\color{white}.} Orig  & Greedy & Reduce %Reduction
        & {\color{white}..} Orig  & Greedy & Reduce %Reduction
        & {\color{white}..} Total & {\color{white}.} Proof  & {\color{white}.} Proof \% \\[2pt]  %Proof/Total \\
      \hline \\[-8pt]
      Datapath 1 & 174  & 90  & 48\%  & 67  & 61  & 9\%  & 37.5 &   2.58 & 7\% \\[2pt]
      Datapath 2 & 561  & 92  & 84\%  &	98  & 46  & 53\% & 34.5 &   2.08 & 6\%\\[2pt]
      Datapath 3 &  14  & 13  & 7\%   &	13  & 12	& 8\%  & 5.13 &   0.49 & 9\% \\[2pt]
      Datapath 4 & 4402 &	202 & 95\%  &	223 &	120 & 46\% & 76.4 &  32.80 & 43\%\\[2pt]
      Datapath 5 & 271  &	95	& 65\%  &	101 &	72	& 29\% & 105  &   0.18 & 0.2\%\\[2pt]
      Datapath 6 & 155  &	83	& 46\%  &	67  &	49	& 27\% & 280  & 168.00 & 60\%\\[2pt]
      %\hline
    \end{tabular}
    \end{center}
    \label{tbl:intel}
  \end{table*}

\subsection{Detailed Analysis}
\label{sec:detailed}

In this section, we perform a more detailed ablation study comparing \egg's
  results using different algorithms.
We implement proof reduction for \egg and the optimal tree width algorithm 
  described in \Cref{sec:optimal}.
The ILP solution is not feasible to run, so we use
  Z3 as a baseline.

\autoref{fig:algorithms-table} summarizes the results.
Z3 and \egg are optimized for different workloads and so
  use different underlying congruence closure algorithms,
  and so produce different proofs.
Using proof reduction, \egg finds slightly shorter proofs than
  Z3.
It also performs better than Z3-style proof reduction implemented in \egg.
Using the greedy algorithm, \egg finds proofs which are even shorter,
  and which are also quite close to proofs of optimal tree size.
The data in \autoref{fig:algorithms-table} consists of the
  \zdagsizevsoptimalnumbenchmarks out of \dagsizenumbenchmarks where Z3 did not time out,
  the same set used in \autoref{fig:dag-size-cdf-optimal}.

While we would ideally use
  the minimal DAG size proofs as
  a baseline in our evaluation,
  we found the ILP formulation
  was infeasible to run on real queries.
However, the $O(n^5)$ \treeopt algorithm,
  which runs in $O(n^3)$ time when the number of congruences is bounded,
  performs well enough to run on all of the examples.
We found that in \percentgreedyequaloptimal
  of these cases,
  the greedy algorithm
  in fact found the proof
  with optimal tree size.
Moreover,
  across all of these benchmarks
  our greedy algorithm's overall
  performance closely tracks
  that of \treeopt,
  showing that the greedy algorithm's
  proof certificates are difficult
  to shrink further.

%Now that the various proof production algorithms have been compared,
%  we turn our attention to the proof tree size measure
%  as a proxy for DAG size.
%We performed experiments
%  to compare the ILP, optimal, and greedy
%  proof length optimization algorithms,
%  shown in \autoref{fig:compare-algs}.
%Unfortunately, the ILP formulation is infeasible to
%  run on real-world examples.
%Instead, we leverage the fact that
%  the ILP algorithm finds the proof with optimal DAG size
%  while the optimal algorithm of \Cref{sec:optimal}
%  finds the proof with optimal tree size.
%\autoref{fig:compare-algs} compares tree sizes to DAG sizes
%  for \egg's proofs, both with and without proof optimization;
%  in both cases, the two measures are correlated,
%  suggesting that optimizing tree size using the optimal algorithm
%  will tend to improve DAG size as well.
%Then, to compare the optimal and greedy algorithms,
%  we ran both on all of our input/output pairs.
%The optimal algorithm found proofs only a bit shorter than the greedy algorithm.
%In \percentgreedyequaloptimal of cases, the greedy algorithm
%  in fact found the proof with optimal tree size.
%These experiments show that the reduction
%  from ILP to optimal to greedy algorithm
%  retains most of the practical benefit
%  of proof length optimization.

\subsection{Case Study}

Typically, proof production is necessary in equality saturation
  to perform translation validation.
In this case, the shorter proofs produced by proof length optimization
  reduce the number of translation validation steps that must be performed
  and thus result in faster end-to-end results.
A practical application that benefits from this reduction is hardware
  optimization performed using \egg by researchers at
  Intel Corporation~\cite{hardware-egraph-optimisation}.
Translation validation is used to ensure that the \egg optimized
  hardware designs are formally equivalent to the input.
Extremely high assurance is needed for hardware designs
  because of the high cost of actual hardware manufacturing.
For \emph{each} step in the tree proof two Register Transfer Level (RTL)
  designs are generated, which are proven to be formally equivalent by
  Synopsys HECTOR technology, an industrial formal equivalence checking tool.
The intermediate steps generate a chain of reasoning proving the equivalence of
  the input and optimized designs, necessary because the tools can
  fail to prove equivalence of significantly transformed designs.
The tree proof is used to ensure that HECTOR can prove each step
  with no user input as it is a simpler check than a DAG proof step.

\begin{figure}
  \begin{tikzpicture}[every matrix/.style={ampersand replacement=\&,column sep=1.5cm,row sep=.4cm}]
  % Position the nodes using a matrix layout
  \matrix{
                                                 \&                                                   \& \node [shape=ellipse, draw=black] (+5ab) {$+$}; \& \node (5ab) {$5a+b$};\\
    \node [shape=ellipse, draw=black] (a) {$a$}; \& \node [shape=ellipse, draw=black] (+2ab) {$+$}; \&  \& \node (4a2b) {$4a+2b$};\\
                                                 \& \node [shape=ellipse, draw=black] (-ab) {$-$};  \& \node [shape=ellipse, draw=black] (+3a3b) {$+$};\&\node (3a3b) {$3a+3b$};\\
    \node [shape=ellipse, draw=black] (b) {$b$}; \& \node [shape=ellipse, draw=black] (+a2b) {$+$}; \& \& \node (2a4b) {$2a+4b$};\\
                                                 \&                                                   \& \node [shape=ellipse, draw=black] (+a5b) {$-$}; \& \node (a5b) {$a+5b$};\\
    };

    \draw[->] (a) -- (+2ab) node [midway,above] {\footnotesize $<<1$};
    \draw[->] (a) -- (-ab);
    \draw[->] (a) -- (+a2b);

    \draw[->] (b) -- (+2ab);
    \draw[->] (b) -- (-ab);
    \draw[->] (b) -- (+a2b) node [midway,below] {\footnotesize$<<1$};
    
    \draw[->] (+2ab) -- (+5ab) node [midway,above=3pt] {\footnotesize$<<1$};
    \draw[->] (+2ab) -- (4a2b) node [midway,above] {\footnotesize$<<1$};
    \draw[->] (+2ab) -- (+3a3b);
    
    \draw[->] (-ab) -- (+5ab);
    \draw[->] (-ab) -- (+a5b);
      
    \draw[->] (+a2b) -- (+3a3b);
    \draw[->] (+a2b) -- (2a4b) node [midway,below] {\footnotesize$<<1$};
    \draw[->] (+a2b) -- (+a5b) node [midway,below=3pt] {\footnotesize$<<1$};
    
    \draw[->] (+5ab) -- (5ab);
    \draw[->] (+3a3b) -- (3a3b);
    \draw[->] (+a5b) -- (a5b);
\end{tikzpicture}
  \caption{Dataflow graph of an optimized multiple constant multiplication circuit design generated by \egg.
    %An accompanying proof to formally prove the equivalence against the input design is also generated by \egg.
    }
  \label{fig:intel_dfg}
\end{figure}
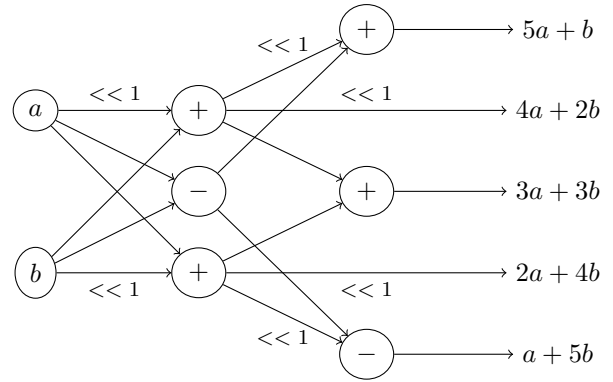

The results of evaluating this paper's greedy optimization algorithm
  on six Intel-tested RTL design benchmarks
  are shown in \autoref{tbl:intel}.
On average, proof lengths decreased by $29\%$,
  with the best case showing a $53\%$ reduction,
  while proof production took only $34$ seconds on average,
  miniscule compared to multi-hour translation validation times.
Moreover, these reductions in proof length resulted
  in shorter translation validation times.
The optimized constant multiplication hardware design descibed in \Cref{fig:intel_dfg} was generated
  by \egg, starting from an initial naive implementation.
Running the complete verification flow for the original and greedy proofs,
  the runtime was reduced from $4.7$ hours to $2.3$ hours.
In more complex examples we expect that days of computation could be saved.
For parameterizable RTL,
  where a design must typically be re-verified for every possible paramterization,
  these gains add up quickly.

\section{Conclusion and Future Work}
\label{sec:conclusion}

% In this paper we explored the problem of
%   finding minimal proofs
%   from first principles.
This paper examined the problem of
  finding minimal congruence proofs
  from first principles.
%Though past work had noted in passing that
%  this problem was NP-hard,
%  justification was limited to
%  personal communication.
Since finding the optimal solution is infeasible,
 we introduced a relaxed metric for proof size
 called \emph{proof tree size},
 and gave an $O(n^5)$ algorithm
 for optimal solutions in that metric.
% Aside from our reduction from
%   the hitting set problem,
%   our formulation led to a
%   polynomial algorithm which
%   finds optimal proof tree size certificates.
While the optimal algorithm
  is too expensive in practice,
  it provides a reasonable
  baseline for small congruence problems,
  and inspired a practical $O(n \log(n))$
  greedy algorithm which generates
  proofs which are \optimaldagsizevsgreedyotherpercentasbig
  as big on average.

We implemented proof generation in the
  \egg equality saturation toolkit,
  making it the first
  equality saturation engine with this capability.
Since
  equality saturation toolkits---unlike SMT solvers---support
  optimization directly,
  this opens the door to
  certifying the results of much
  recent work in optimization and program
  synthesis~\cite{egg,ruler,tensat,herbie,diospyros,spores,szalinski}.

Looking forward,
  we are especially eager
  for the community to explore
  more applications of proof certificates
  in congruence closure procedures.
For example,
  it should be possible to use proofs
  to tune rewrite rule application
  schedules in e-matching,
  improve debugging of subtle
  equality saturation issues,
  and enable equality-saturation-based
  ``hammer'' tactics in proof assistants.
It may also be possible to further
  improve on the greedy proof generation
  algorithm with better heuristics
  for estimating proof sizes, or
  to enable more efficient prover
  state serialization via smaller proofs.

\section{Acknowledgements}

This work was supported by the Applications Driving Architectures (ADA) Research Center, a JUMP Center co-sponsored by SRC and DARPA and supported by the National Science Foundation under Grant No. 1749570.

\bibliographystyle{IEEEtran}
\bibliography{IEEEabrv,references}

\end{document}